\documentclass[pra,aps,floatfix,twocolumn,longbibliography]{revtex4-1}
\usepackage{amsmath}
\usepackage{amsthm}
\usepackage{mathrsfs}
\usepackage{graphicx}
\usepackage{color}

\DeclareMathOperator{\Tr}{tr}
\newcommand{\dee}{\mathrm{d}}

\newcommand{\ket}[1]{|#1\rangle}

\newcommand{\bracket}[2]{\langle #1|#2\rangle}

\newcommand{\TO}{\mathcal{T}}

\begin{document}

\title{Theory of high-efficiency sum-frequency generation for single-photon waveform conversion}

\author{John~M.~Donohue}
\email[]{jdonohue@uwaterloo.ca}
\address{Institute for Quantum Computing and Department of Physics \&
Astronomy, University of Waterloo, Waterloo, Ontario, Canada N2L 3G1}

\author{Michael~D.~Mazurek}
\address{Institute for Quantum Computing and Department of Physics \&
Astronomy, University of Waterloo, Waterloo, Ontario, Canada N2L 3G1}

\author{Kevin~J.~Resch}
\address{Institute for Quantum Computing and Department of Physics \&
Astronomy, University of Waterloo, Waterloo, Ontario, Canada N2L 3G1}

\begin{abstract}
\noindent The optimal properties for single photons may vary drastically between different quantum technologies. Along with central frequency conversion, control over photonic temporal waveforms will be paramount to the effective coupling of different quantum systems and efficient distribution of quantum information. Through the application of pulse shaping and the nonlinear optical process of sum-frequency generation, we examine a framework for manipulation of single-photon waveforms.  We use a non-perturbative treatment to determine the parameter regime in which both high-efficiency and high-fidelity conversion may be achieved for Gaussian waveforms and study the effect such conversion techniques have on energy-time entanglement.  Additionally, we prove that aberrations due to time ordering are negligible when the phasematching is nonrestrictive over the input bandwidths.  Our calculations show that ideal quantum optical waveform conversion and quantum time lensing may be fully realized using these techniques.
\end{abstract}

\maketitle

\section{Introduction}

Single photons are the natural choice for many quantum technologies
as they are an ideal carrier of quantum information for communication protocols
and coupling quantum nodes~\cite{knill2001scheme,duan2001long}. To form an effective interface between two quantum systems, it is important that the photon properties, such as the spectrum or spatial mode, match those of the receiver.  Ensuring compatibility will, in general, necessitate adapting properties of the source photon to match those of the receiver using waveform manipulation methods.  Constraints imposed by the no-cloning theorem~\cite{wootters1982single} forbid direct amplification or detect-and-resend approaches, creating a need for highly efficient low-noise quantum waveform conversion methods.

\begin{figure}[t]
  \begin{center}
       \includegraphics[width=1\columnwidth]{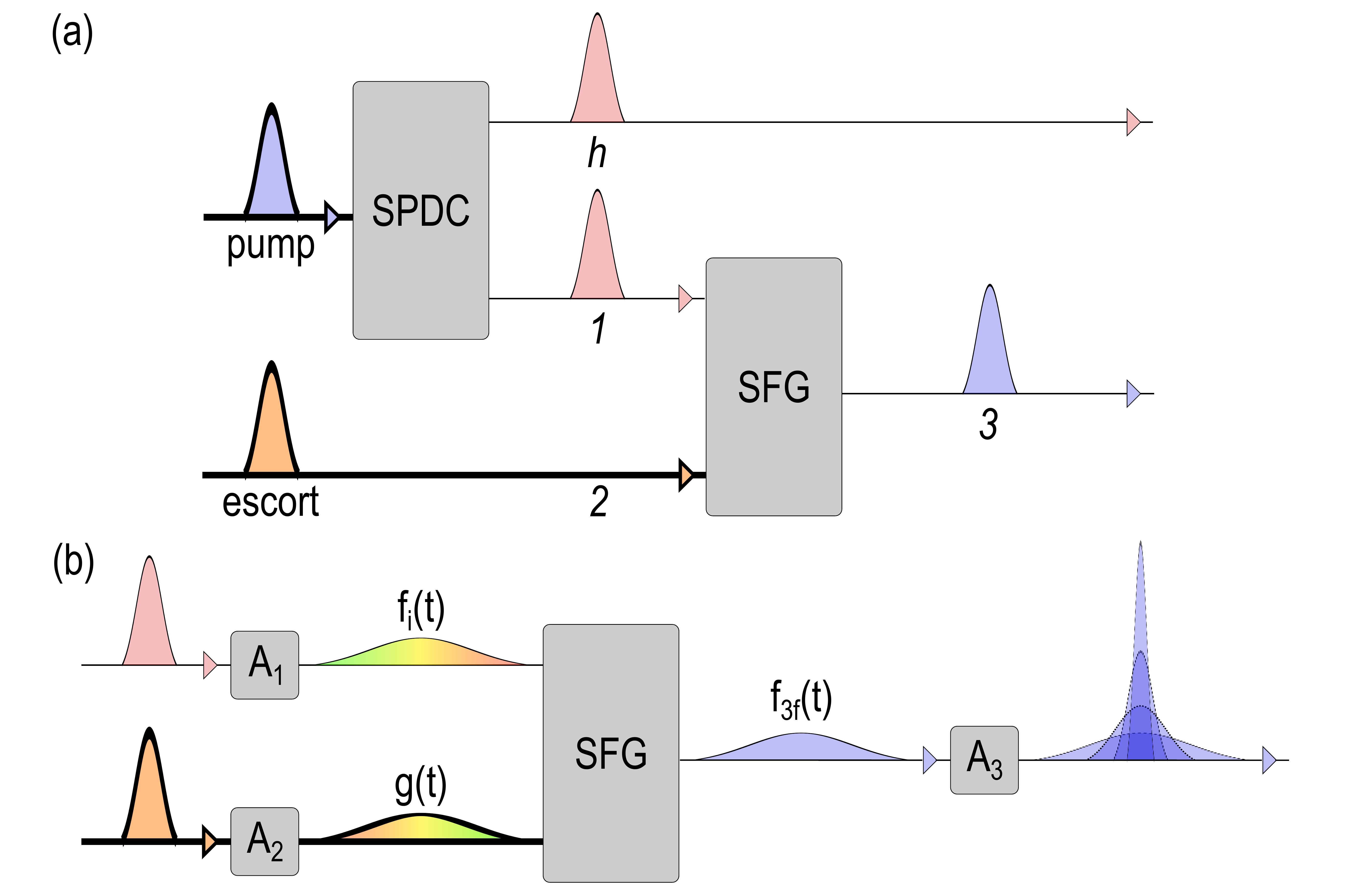}
  \end{center}
 \caption{\textbf{Sum-frequency generation for optical waveform manipulation.} (a) A pair of photons may be created through, for example, spontaneous parametric downconversion, with a signal photon in mode 1 and its herald in mode $h$. The signal photon is then mixed with a strong escort pulse in mode 2 to produce an upconverted signal photon in mode 3. (b) Temporal waveforms may be customized by applying dispersion (represented by the chirp parameter $A_i$) to the input photon and escort pulse before SFG. A third chirp applied to the upconverted signal allows for complete temporal magnification. Bandwidth compression can be achieved in this scenario as shown, with equal-and-opposite chirps applied to the input photon and escort pulse and no third chirp required. Bold lines represent strong pulses and thin lines represent single photons.}\label{concept}
\end{figure}

The temporal waveforms of single photons are of particular importance in quantum optics and quantum information science. Photon pairs produced through spontaneous parametric downconversion (SPDC) have controllable energy-time entanglement dependent on the pump and crystal properties~\cite{grice1997spectral,grice2001eliminating,kim2002generation,mosley2008heralded,bennink2010optimal}. Quantum information may be encoded as a superposition of discretized time~\cite{brendel1999pulsed} or frequency~\cite{olislager2010frequency} bins and multiplexed in either case to increase the rate of information transmission~\cite{jiang2006highly,christ2012exponentially,herbauts2013demonstration,donohue2014ultrafast}.  Control over this degree of freedom is necessary for coupling to quantum memories~\cite{hosseini2011high,england2015storage}, quantum frequency conversion~\cite{kumar1990quantum,mckinstrie2005translation,mcguinness2010quantum,mcguinness2011theory}, temporal mode selection~\cite{eckstein2011quantum,reddy2013temporal,kowligy2014quantum}, and quantum measurement~\cite{donohue13}.  In order for general control over temporal waveforms, it is necessary that the waveform conversion methods remain effective on the ultrafast timescale~\cite{kielpinski,lavoie13comp,agha2014spectral}.

Sum-frequency generation (SFG) and pulse shaping are powerful tools for manipulating photonic temporal waveforms~\cite{raymer2012manipulating}. In classical optics, these processes have been employed to great success in constructing \emph{time lenses}, which can compress or stretch complex waveforms~\cite{kolner1989timelens,kolner1994space,bennett1994temporal,bennett1999upconversion,salem2008optical,foster2009ultrafast,walmsley2009characterization}. In the quantum regime, SFG has been employed to convert telecom band single photons to visible wavelengths for more efficient detection~\cite{kwiat_upconversion,langrock2005highly,vandevender2007quantum,Zaske2012,ates2012two}, and dispersion-controlled SFG has been used for shaping single photons from quantum dot~\cite{rakher2010quantum,rakher2011simultaneous,agha2014spectral} and downconversion sources~\cite{lavoie13comp,donohue13,donohue2014ultrafast}.  This process has been experimentally shown to maintain energy-time entanglement~\cite{tanzilli2005photonic}.  Some applications of waveform manipulation have engineered crystals with specific phasematching functions to customize the output~\cite{arbore1997engineerable,hum2007quasi,eckstein2011quantum}, while others have focused on manipulating the spectrum of the strong laser pulse~\cite{rakher2011simultaneous,kielpinski,lavoie13comp,donohue13,agha2014spectral,donohue2014ultrafast,kowligy2014quantum}.  We restrict our attention to the latter applications, as spectral manipulations to a laser pulse or single photon may be performed with flexible pulse shaping techniques~\cite{weiner2000femtosecond,lukens2013demonstration} while manipulations to the phasematching function generally require specialized crystal engineering.

In recent experiments studying sum-frequency generation between shaped single photons and strong classical fields, a first-order perturbative theory was sufficient to explain the results as the conversion efficiency was low.  However, it is important that these techniques remain effective in the high-efficiency regime.  In this work, we develop a quantum treatment of an idealized sum-frequency generation process between a single photon and a strong classical pulse to address this issue. Our treatment is necessary to enable practical bandwidth compression and time lensing for the quantum domain, as aberrations occurring at high efficiency could greatly degrade the quality of the signal. The paper is structured as follows. In Section II, we derive the quantum waveform resulting from the SFG interaction. We also justify an approach based on the Taylor series expansion of the unitary transformation by showing that the corrections arising from a more complete Dyson or Magnus series treatment vanish in the limiting case where phasematching is nonrestrictive.  In Section III, we apply this result to a model heralded single photon waveform. In Section IV, we review the special cases of time lensing, time-to-frequency conversion, and bandwidth compression as examples of the flexibility of dispersion-based waveform shaping. In Section V, we discuss the effect of waveform conversion on half of an energy-time entangled photon pair.

\section{SFG waveform conversion}

We consider the scenario depicted in Fig.~\ref{concept}(a).  A photon pair is created in modes $1$ and $h$, with the photon in mode $h$ acting as a herald.  These photons may be energy-time entangled and could originate from a process such as spontaneous parametric downconversion.  We model the state of the two photons as~\cite{grice2001eliminating,kim2002generation,resch2009chirped,branczyk2010non} \begin{align}\ket{\psi_i(t)}=&\frac{1}{2\pi} \iint\dee\omega_1\dee\omega_h F_i(\omega_1,\omega_h)\ket{\omega_1}_1\ket{\omega_h}_h,\label{stateinit}\end{align} where $\ket{\omega_j}=\hat{a}_{\omega_j}^{\dag}e^{i\omega_jt}\ket{0}$ is a single-photon of frequency $\omega_j$ in a single spatial mode. For simplicity, we will assume that all limits of integration extend to infinity for the remainder of this discussion.

The single photon in mode $1$ then interacts with a strong escort pulse in mode $2$ through sum-frequency generation to produce upconverted light in mode $3$.  We model this interaction using a unitary transformation which we will discuss briefly here (see Appendix A for details).  The SFG material is assumed to be a $\chi^{(2)}$ medium with a fast nonlinearity and phasematching function ${\Phi(\omega_1,\omega_2,\omega_3)}$, and the input fields are assumed to be plane waves. We treat the escort pulse in mode 2 as a classical non-depleted field with normalized spectrum $G(\omega_2)$.  The strength of the interaction is defined by the \emph{absolute coupling constant} $\gamma$, which is proportional to the electric field amplitude of the escort pulse, the crystal length, and the strength of the nonlinearity of the material, with units such that ${\int\dee\omega_2\,\gamma^2\,|G(\omega_2)|^2}$ is dimensionless.  In this limit, the interaction Hamiltonian may be expressed as~\cite{hillery1984quantization,grice1997spectral,yang2008spontaneous} \begin{widetext}\begin{equation}\hat{H}_I(t)=\,\frac{\hbar\gamma}{2\pi}\iiint\dee\omega_1\dee\omega_2\dee\omega_3\,\left[\Phi(\omega_1,\omega_2,\omega_3)G(\omega_2)e^{-i(\omega_1+\omega_2-\omega_3)t}\hat{a}_{\omega_1}\hat{c}^\dag_{\omega_3}+\mathrm{h.c.}\right],\label{mainham}\end{equation}\end{widetext} where the operators $\hat{a}$ and $\hat{c}$ correspond to modes $1$ and $3$, respectively.

The unitary transformation describing time evolution is written as \begin{equation}\hat{U}=\TO\exp\left[-\frac{i}{\hbar}\int_{t_0}^{t_f}\dee t\hat{H}_I(t)\right]\label{expunitarymain},\end{equation} where $\TO$ is the \emph{time-ordering operator}, which ensures proper ordering of events for Hamiltonians that do not commute with themselves at different times.  This operator prevents a straightforward Taylor expansion of the unitary transformation, necessitating a treatment through the Dyson or Magnus expansion, as has been shown in previous work in quantum optics~\cite{branczyk2010non,branczyk2011time,christ13theory,quesada2014effects,quesada2014time}.  Here we simplify calculations by assuming that the phasematching function $\Phi(\omega_1,\omega_2,\omega_3)$ has a wide acceptance bandwidth and set it to one across the frequencies of interest.  In this limit, which corresponds to a process where the fields in each of the three modes co-propagate without temporal walkoff or spread, we find that the time-ordering corrections to the Taylor expansion vanish.  We prove this by showing that, in this limit, the commutator of the Hamiltonian with itself at different times is zero, i.e. \begin{equation}\left[\hat{H}_I(t_2),\hat{H}_I(t_1)\right]=0.\label{notimeorder}\end{equation}  This approximation is expected to be valid for thin nonlinear crystals; it could also be achieved using nonlinear materials with a flat (top-hat) phasematching function much broader than the escort bandwidth~\cite{suhara1990theoretical,dosseva2014shaping}.

To show the commutativity given by Eq.~\eqref{notimeorder}, we define the following:  \begin{align}\Xi(t)=&\,\hbar\gamma\int\dee\omega_2\,G(\omega_2)e^{-i\omega_2t}, \\\hat{a}(t)=&\,\frac{1}{\sqrt{2\pi}}\int\dee\omega_1\,\hat{a}_{\omega_1}e^{-i\omega_1t},\label{atdef} \\\hat{c}(t)=&\,\frac{1}{\sqrt{2\pi}}\int\dee\omega_3\,\hat{c}_{\omega_3}e^{-i\omega_3t}.\label{ctdef}\end{align} We then substitute these functions and operators into the Hamiltonian of Eq.~\eqref{mainham} along with the assumption $\Phi(\omega_1,\omega_2,\omega_3)=1$ to find the simplified Hamiltonian of \begin{equation}\hat{H}_I(t)=\Xi(t)\hat{a}(t)\hat{c}^\dag(t)+\Xi^*(t)\hat{a}^\dag(t)\hat{c}(t).\end{equation}  Note that this factorization of the Hamiltonian is not generally possible for a non-constant phasematching function.  It is an interesting question whether there exist other symmetries in the phasematching function that also lead to the relation of Eq.~\eqref{notimeorder}. The operators $\hat{a}(t)$ and $\hat{c}(t)$ as defined in Eqs.~\eqref{atdef} and \eqref{ctdef} can be shown to obey the commutation relation ${\left[\hat{a}(t_2),\hat{a}^\dag(t_1)\right]=\delta(t_2-t_1)}$ from the frequency-domain relation ${\left[\hat{a}_{\omega_2},\hat{a}^\dag_{\omega_1}\right]=\delta(\omega_2-\omega_1)}$.  Using these commutation relations to normal-order $\hat{H}_I(t_2)\hat{H}_I(t_1)$ and $\hat{H}_I(t_1)\hat{H}_I(t_2)$, it can be seen that Eq.~\eqref{notimeorder} is satisfied, and thus the Taylor series is sufficient to obtain the SFG unitary transformation.  This statement holds independently of the input state $\ket{\psi_i(t)}$ and escort spectrum $G(\omega_2)$.  Recently, Quesada and Sipe proved from the Magnus expansion that the second- and third-order corrections to the Taylor expansion vanished for broadly phasematched processes in both SPDC and SFG~\cite{quesada2014effects}; based on our result, we conclude that the corrections to the Taylor series vanish for all orders for both the Magnus and Dyson expansion.

We rewrite the sum-frequency generation unitary transformation as ${\hat{U}=\sum_{k=0}^{\infty}\frac{(i\gamma)^k}{k!}\hat{U}_k}$, where  \begin{align}\hat{U}_k=\iint\dee\omega_1\dee\omega_3\,\left[\hat{a}_{\omega_1}\hat{c}_{\omega_3}^{\dag} G(\omega_3-\omega_1)+\mathrm{h.c}.\right]^k.\label{unitary4}\end{align}  Applying this unitary transformation to our initial state from Eq.~\eqref{stateinit} gives a final state with even-order terms in the input mode 1 and odd-order terms in the upconverted mode 3.  Each term in the series may be found through a recursion relation. The representation of the state has a closed-form solution in the time domain, which can be found using the convolution theorem. Details on this calculation can be found in Appendix B; we will state the result here. The full spectral waveforms of the remaining photon amplitude in mode 1 is given by ${F_{1f}(\omega_1,\omega_h)=\sum_{k=0}^\infty F^{(2k)}(\omega_1,\omega_h)}$ and the upconverted signal in mode 3 is ${F_{3f}(\omega_3,\omega_h)=\sum_{k=0}^\infty F^{(2k+1)}(\omega_3,\omega_h)}$.
The initial joint spectral amplitude and escort spectrum can be expressed in the time domain as $f_i(t,t_h)$ and $g(t)$, respectively, using the Fourier transform ${f(t)=\frac{1}{\sqrt{2\pi}}\int\dee\omega\,F(\omega)e^{i\omega t}}$.  The final two-photon temporal waveforms for the photons in modes $h$ and either $1$ or $3$ can be written in the time domain as
\begin{align}f_{1f}(t,t_h)&=f_i(t,t_h)\cos\left[\sqrt{2\pi}\gamma|g(t)|\right],\label{f1f}\\
f_{3f}(t,t_h)&=f_i(t,t_h)\frac{g(t)}{|g(t)|}\sin\left[\sqrt{2\pi}\gamma|g(t)|\right].\label{f3f}\end{align}  These equations are consistent with the result for the exactly co-propagating signals regime in~\cite{reddy2013temporal}.

\section{Waveform conversion of a single photon from a model energy-time entangled pair}

In order to characterize the effectiveness and efficiency of these processes, we model our input state as part of a photon pair produced through SPDC in a broadly phasematched crystal followed by bandpass filters, which we write as \begin{align}F_i(\omega_1,\omega_h)=&\frac{(S^2+\sigma_1^2+\sigma_h^2)^{\frac{1}{4}}}{\sqrt{2\pi S\sigma_1\sigma_h}}e^{iA_1(\omega_1-\omega_{01})^2}\label{inputspec} \\&e^{-\frac{(\omega_1-\omega_{01})^2}{4\sigma_1^2}-\frac{(\omega_h-\omega_{0h})^2}{4\sigma_h^2}-\frac{(\omega_1+\omega_h-\omega_{01}-\omega_{0h})^2}{4S^2}}\nonumber,\end{align} where $\sigma_1$ and $\sigma_h$ are the bandwidths of spectral filters centred at frequencies $\omega_{01}$ and $\omega_{0h}$, respectively, and $S$ is the bandwidth of the pump beam~\cite{kim2002generation,resch2009chirped,branczyk2010optimized}.  The chirp parameter $A_1$ determines the strength of the group velocity dispersion applied to the signal photon.

Our escort beam is described by a normalized Gaussian spectrum with group velocity dispersion as \begin{equation} G(\omega_2)=\frac{1}{(2\pi\sigma_2^2)^{\frac{1}{4}}}e^{-\frac{(\omega_2-\omega_{02})^2}{4\sigma_2^2}}e^{i\omega_2\tau}e^{iA_2(\omega_2-\omega_{02})^2},\label{escortspec}\end{equation} where $\sigma_2$ is the bandwidth of the escort about central frequency $\omega_{02}$, $A_2$ applied dispersion, and $\tau$ a time delay relative to the signal photon.

\begin{figure*}
  \begin{center}
    \includegraphics[width=2\columnwidth]{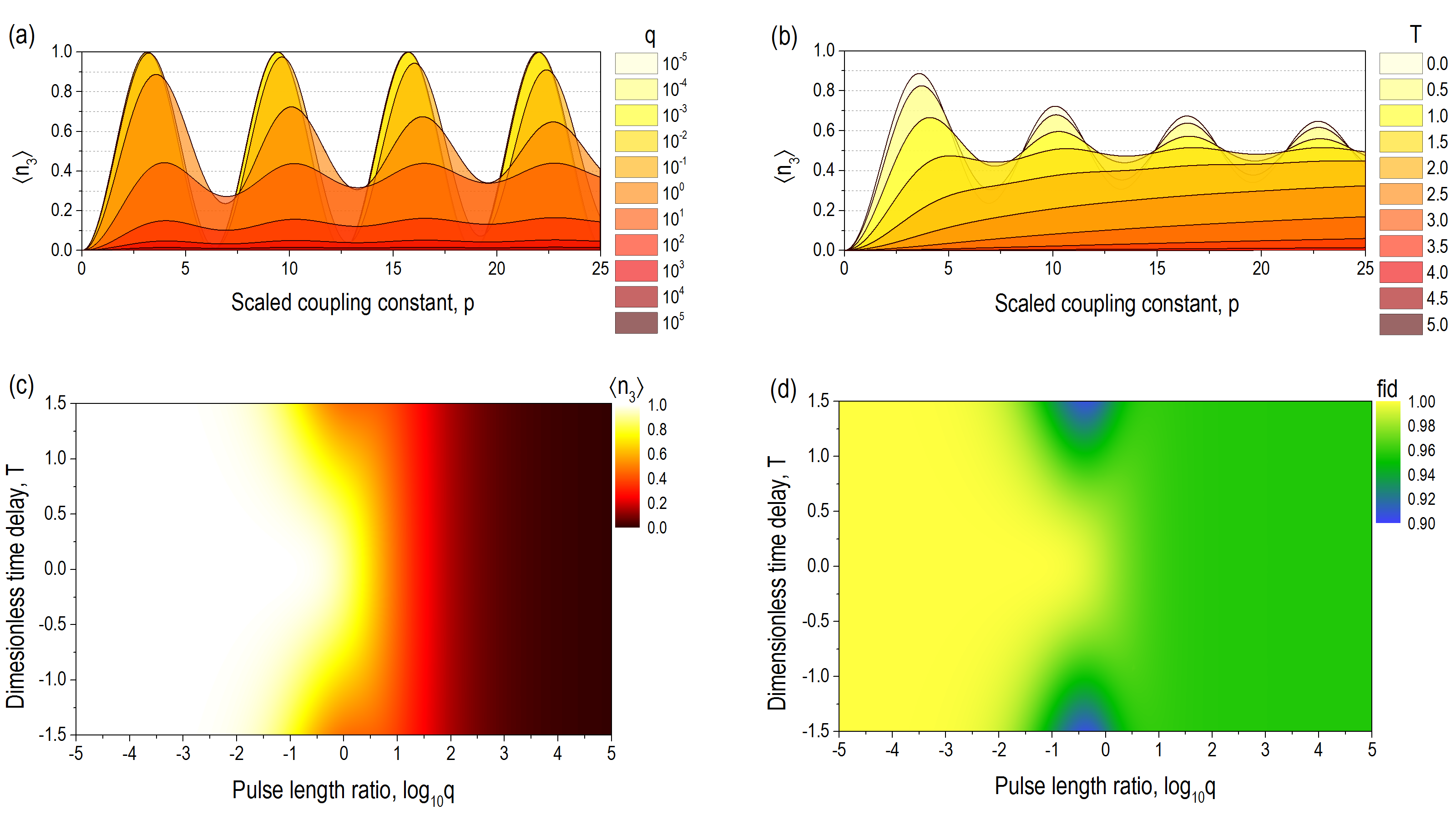}
  \end{center}
 \caption{\textbf{Single-photon upconversion efficiency and fidelity.} (a) The probability of successful upconversion $\langle\hat{n}_3\rangle$ is shown as a function of the scaled coupling constant $p$ for various pulse width ratios $q$, with the dimensionless time delay $T$ held constant. As $p$ increases in the regime where the escort is much broader in time than the input photon (low $q$), the efficiency of upconversion follows a sine pattern, reaching unit efficiency at $p=\pi$. In the regime where the escort is much narrower in time than the photon, high upconversion efficiency is not achievable. (b) Here we show the probability as a function of $p$ for various dimensionless time delays $T$, with the pulse width ratio $q$ held at one.  As the time delay is increased, the pulses cease to overlap well and the maximum efficiency is seen to drop. Notably, the peak efficiency is no longer well defined past $|T|\approx1.5$, as the first local maximum in efficiency is no longer the global maximum. (c) The maximum possible efficiency is numerically calculated as a function of $q$ and $T$, with the optimal $p$ estimated as the first peak of a fourth-order expansion of Eq.~\eqref{SumGen}.  In the low-$q$ regime, the efficiency is also robust against time delays $T$. (d) By numerically calculating the fidelity of the temporal waveform at the estimated optimal efficiency with that expected from first-order perturbation theory via Eq.~\eqref{fid}, we see that the first-order approximation is an excellent description in the low-$q$ regime.  Time delays disturb the symmetry of the system and further reduce the fidelity.}\label{series}
\end{figure*}

We may now calculate the probability of successfully converting the single photon from mode 1 to mode 3, by finding the expectation value of the number of photons in mode 3, $\langle\hat{n}_3\rangle$.  We Fourier transform the spectra described by Eqs.~\eqref{inputspec} and \eqref{escortspec} and substitute them into Eq.~\eqref{f3f} to find the two-photon waveform in modes $3$ and $h$.  We can then calculate $\langle\hat{n}_3\rangle$ by integrating the square of this amplitude over all time, \begin{align}\langle\hat{n}_3\rangle=&\int\dee t\dee t_h |f_{3f}(t,t_h)|^2\nonumber\\=&\frac{1}{2}\sum_{k=1}^\infty\frac{(-1)^{k-1}}{(2k)!}\frac{e^\frac{-kT^2}{1+qk}}{\sqrt{1+qk}}p^{2k}\label{SumGen}\end{align} where we have made the following substitutions: \begin{align}p=&\,2(8\pi)^{\frac{1}{4}}\left(\frac{\sigma_2^2}{1+16A_2^2\sigma_2^4}\right)^\frac{1}{4}\gamma\,,\label{SumGenp}\\ T=&\,\frac{\sqrt{2}\sigma_2\tau}{\sqrt{1+16A_2^2\sigma_2^4}}\,,\label{SumGenT} \\q=&\,\frac{\sigma_2^2}{\sigma_1^2}\frac{\left[1+\frac{\sigma_1^2}{S^2}+\frac{16A_1^2\sigma_1^4(S^2+\sigma_h^2)}{S^2+\sigma_1^2+\sigma_h^2}\right]}{1+16A_2^2\sigma_2^4}\,.\label{SumGenq}\end{align} The \emph{scaled coupling constant} $p$ is defined such that $p^2$ is proportional to the peak power of the escort pulse, which may be seen by noting that $\gamma^2$ is proportional to the number of photons in the pulse and its stretched temporal duration is given by $\Delta{t}^2\propto\frac{1+16A_2^2\sigma_2^4}{\sigma_2^2}$.  The \emph{dimensionless time delay} $T$ corresponds to a relative time delay between the escort pulse and the photon normalized to the temporal width of the escort pulse. Finally, the \emph{pulse length ratio} $q$ is the ratio of temporal widths for the escort pulse and the input photon, with a low value implying that the single photon is much shorter in duration than the escort pulse.  This series is provably convergent for any value of $p$, $q$, or $T$ through the Cauchy-Hadamard theorem~\cite{CauchyHadamard}.

The three parameters of Eqs.~(\ref{SumGenp}-\ref{SumGenq}) characterize the important figures of merit for the conversion process. In particular, the pulse length ratio $q$ describes the potential efficiency of a given sum-frequency process. In the low-$q$ limit, where $1+qk\approx1$ for all $k$ with appreciable contributions to Eq.~\eqref{SumGen}, it is seen that \begin{equation}\lim_{q\rightarrow0}\langle\hat{n}_3\rangle=\sin^2\left(\frac{1}{2}e^{-\frac{T^2}{2}}p\right),\end{equation} as one would find by treating the escort pulse as monochromatic~\cite{BoydNonlinearOptics}.  In this limit, perfect upconversion efficiency ($\langle\hat{n}_3\rangle=1$) is achievable with sufficient escort power. In the high-$q$ limit, Eq.~\eqref{SumGen} does not readily present a closed-form solution and must be studied numerically.  Fig.~\ref{series}(a) shows numerical calculations of $\langle\hat{n}_3\rangle$ as a function of the scaled coupling constant $p$ for a wide range of $q$ values with zero time delay ($T=0$).  It is apparent that high SFG efficiency may only be achieved for low values of the pulse length ratio, $q\lesssim1$.  Fig.~\ref{series}(b) shows $\langle\hat{n}_3\rangle$ as a function of $p$ as the dimensionless time delay $T$ is varied, for equal escort and photon pulse lengths ($q=1$).  These calculations show that adding a time delay will also reduce the peak efficiency, as expected due to the decrease in overlap between the escort pulse and single photon.

To find the optimal conversion efficiency, we aim to find the maximal $\langle\hat{n}_3\rangle$ for any given $q$ and $T$.  We estimate the optimum $p$ value as the first zero of the derivative of Eq.~\eqref{SumGen} with respect to $p$, corresponding to the first efficiency peak in Fig.~\ref{series}(a-b).  As this derivative is not in closed form, we truncate the sum of Eq.~\eqref{SumGen} after four terms to obtain an approximate maximum; this works well for oscillatory solutions, but may underestimate the optimal $p$ for large values of $|T|$ as no well-defined peak efficiencies are present.  In Fig.~\ref{series}(c), we show the optimal efficiency using this method as a function of $q$ and $T$.  This optimal efficiency is nearly unity for small pulse length ratios $q\lesssim1$ and very robust against time delays for $q\ll1$.  However, note that higher escort power is required to reach the optimal efficiency as the time delay moves away from zero.

A first-order perturbative approach, as used in previous works, always predicts a Gaussian sum-frequency photon given a Gaussian input photon and escort pulse.  This is an ideal target photon for many applications, and it is important to determine how well this relatively simple prediction describes the result expected at high efficiency.  By defining $\ket{\psi^{(1)}}$ to be the photonic waveform found through first-order perturbation theory, with a temporal waveform $f_{3f}^{(1)}(t,t_h)$ found by expanding Eq.~\eqref{f3f} to first order in $\gamma$, we can calculate how well our total photonic waveform overlaps with the first-order description through the quantum state fidelity, and thus determine the validity of first-order approximations. The fidelity is defined for pure states as \begin{equation}\left|\bracket{\psi^{(1)}}{\psi}\right|^2=\frac{\left|\iint\dee t\dee t_h f_{3f}^*(t,t_h)f_{3f}^{(1)}(t,t_h)\right|^2}{\langle\hat{n}_3^{(1)}\rangle\langle\hat{n}_3\rangle}, \label{fid}\end{equation} where we normalize by dividing by both $\langle\hat{n}_3\rangle$ and ${\langle\hat{n}_3^{(1)}\rangle=\iint\dee t\dee t_h \left|f_{3f}^{(1)}(t,t_h)\right|^2}$ as we are primarily concerned with the shape of the temporal waveforms.  We numerically calculate this fidelity at the optimal efficiency as a function of $q$ and $T$, with the results shown in Fig.~\ref{series}(d). It is seen that the first-order approximation describes the high-efficiency waveform well in the low-$q$ regime, but is less accurate when $q$ is large; however, the fidelity is numerically always above 0.95 as long as there is no relative time delay between the photon and the escort ($T=0$). When a time delay is introduced, the fidelity dips as a function of $T$ for moderate values of $q$, as considerable pulse reshaping occurs.

\section{Applications of quantum optical waveform manipulation}

\subsection{Time lensing}

The space-momentum and energy-time properties of light can be described by similar mathematical structures, allowing analogous techniques for controlling these properties~\cite{kolner1994space}. For example, in the paraxial and relatively narrowband limits respectively, beam diffraction in free space, the action of a lens, and group velocity dispersion are mathematically described by quadratic phases in transverse momentum, space, and frequency, respectively.  The additional ability to produce quadratic temporal phases, known as \emph{times lenses}, in conjunction with group velocity dispersion enables temporal magnification of optical pulses in the same way that the combination of lenses and free space propagation provide the ingredients for spatial magnification~\cite{kolner1989timelens,kolner1994space}.  Group velocity dispersion may be applied to single photons using pulse shapers and dispersive waveguides in the same fashion as strong laser pulses~\cite{lukens2013demonstration,lavoie13comp}, but the quadratic temporal phase needed for the time lens is more challenging.  The nonlinear optical process of self-phase modulation can serve as a time lens for strong laser pulses, but is ineffective for single photons~\cite{garrison2008quantum}.  Microwave electronic phase modulators have also been used as time lenses~\cite{kolner1994space}, but are only effective for limited pulse bandwidths~\cite{walmsley2009characterization}.

Instead of applying the temporal phase directly, it is possible to apply dispersion to a strong escort pulse and induce a temporal phase through SFG~\cite{bennett1994temporal,bennett1999upconversion}. To see how the temporal phase is imparted to the single photon in SFG, we examine Eq.~\eqref{f3f} in the limit where the overall temporal width of the strong escort pulse is much greater than the temporal width of the input single photon ($q\ll1$); in this limit, we make the approximation that ${\sin\left[\sqrt{2\pi}\gamma|g(t)|\right]\approx\sin\left[\sqrt{2\pi}\gamma\right]}$. This limit is fortuitously identical to the required limit for high conversion efficiency, as seen in the previous section.  The output temporal waveform of Eq.~\eqref{f3f} is then modified relative to the input $f_i(t,t_h)$ by the chirped escort pulse $g(t)$ given by the Fourier transform of Eq.~\eqref{escortspec}, with an imparted quadratic temporal phase $\phi(t)=Bt^2$, where the coefficient is \begin{equation}B=\frac{-4A_2\sigma_2^4}{1+16A_2^2\sigma_2^4}.\end{equation}

With the addition of a third dispersive element imparting chirp $A_3$ on the output, as shown in Fig.~\ref{concept}(b), a temporal waveform may be imaged with magnification ${M=-(A_1/A_3)}$ so long as the condition \begin{equation}\frac{1}{2A_1}+\frac{1}{2A_3}=2B=\frac{-8A_2\sigma_2^4}{1+16A_2^2\sigma_2^4}\stackrel{LCL}{\approx}-\frac{1}{2A_2}\label{tlens}\end{equation} is met~\cite{kolner1989timelens,walmsley2009characterization}. The limit for the approximation in Eq.~\eqref{tlens} is the \emph{large-chirp limit} (LCL) for the strong escort pulse such that ${A_2^2\sigma_2^4\gg1}$. This leads to a convenient representation of temporal imaging that has a form identical to the familiar thin lens equation from ray optics, with the chirps $A_1$ and $A_3$ applied to the signal acting as object and image distance and $-A_2$ acting as the focal length.

Under slightly different conditions, the same configuration may be used as a \emph{time-to-frequency converter}~\cite{kauffman1994time,arons1997high,walmsley2009characterization}, where the time profile is mapped to the spectrum and vice versa in direct analogy to a spatial Fourier transform. Once again, a chirp on both the input and output photon is required, this time with equal chirp parameters $A_1=A_3$. The time lens required should impart a quadratic phase $Bt^2$ such that \begin{equation}\frac{1}{4A_1}=B\stackrel{LCL}{\approx}-\frac{1}{4A_2},\end{equation} or $A_1=-A_2=A_3$ in the large-chirp limit.

\subsection{Bandwidth compression}

\begin{figure}[t]
  \begin{center}
    \includegraphics[width=1\columnwidth]{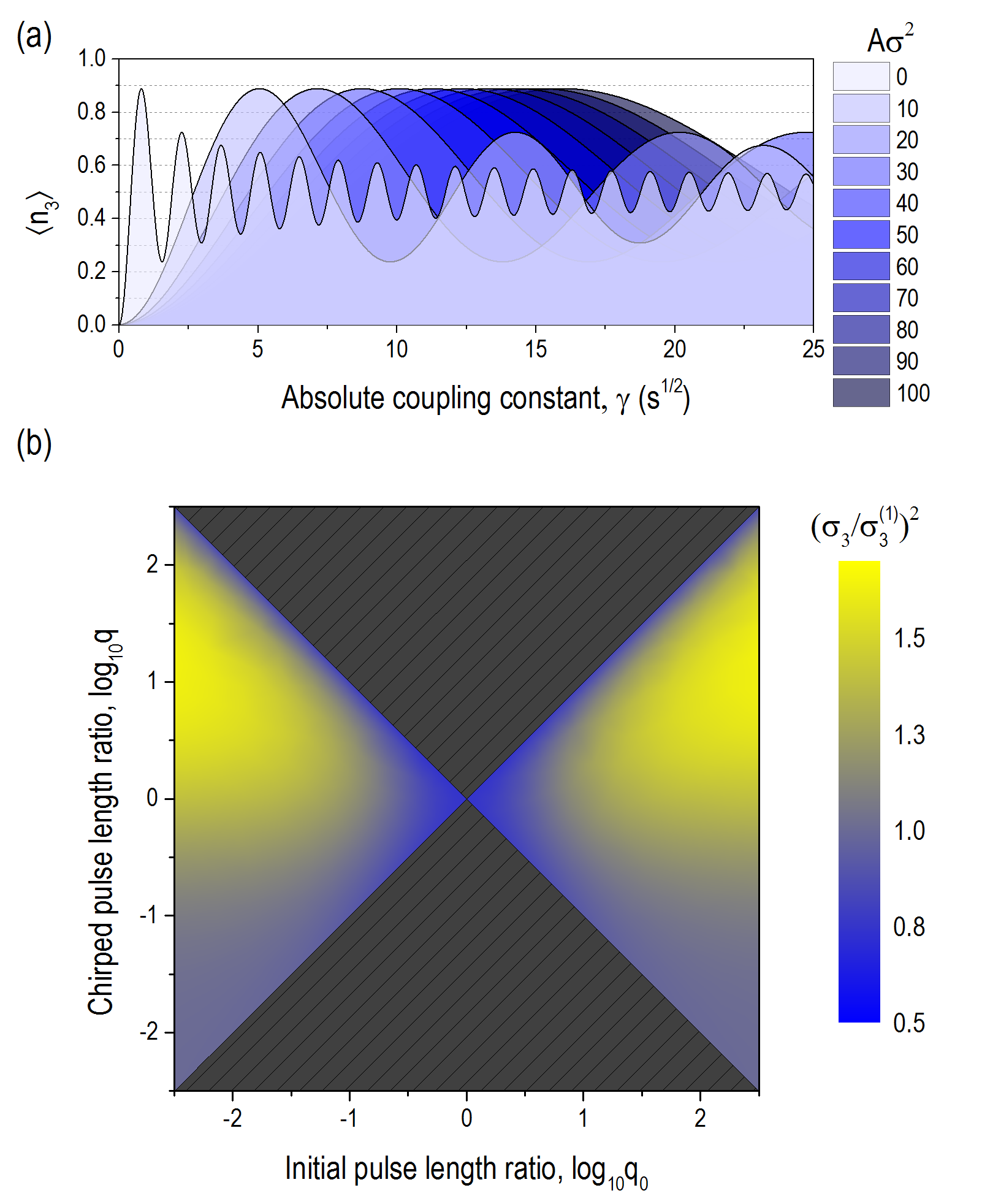}
  \end{center}
 \caption{\textbf{Effectiveness and efficiency of bandwidth compression.} (a) The success probability $\langle\hat{n}_3\rangle$ of bandwidth compression is shown as a function of the absolute coupling constant $\gamma$, with $\tau=0$, $\sigma_1=\sigma_2=\sigma$, $A_1=-A_2=A$, and $A_3=0$. As the chirp applied is increased, the compression achieved is stronger at the expense of peak power in the escort pulse; however, while more power is required to achieve optimal efficiency, the potential peak efficiency is constant. (b) In the regime where the pulse length ratio $q$ is low, the spectral width of the upconverted signal $\sigma_3$ is seen to be identical to the width $\sigma_3^{(1)}$ expected from a first-order approximation regardless of the input bandwidths. However, as $q$ grows, the ratio of the two widths is seen to depend on the pulse length ratio before chirp, $q_0$. The lined region is algebraically inaccessible for real-valued chirp parameters $A_1=-A_2$, with the $q=q_0$ line corresponding to zero applied chirp and the $q=\frac{1}{q_0}$ line corresponding to the large-chirp limit.}\label{bc}
\end{figure}

In many photonic applications, it is advantageous to have control over the spectral bandwidth of a single photon. Chirped-pulse upconversion may be used for \emph{bandwidth compression} with equal-and-opposite dispersion on the single photon and escort pulse ($A_1=-A_2$), as has been demonstrated experimentally for single photons in the $q\approx1$ and low-efficiency regime~\cite{lavoie13comp,donohue13}.  Similar schemes have seen classical applications in dispersion-canceled imaging~\cite{kaltenbaek2008quantum,lavoie2009quantum,mazurek2013dispersion}. It has been theoretically shown that entanglement is detrimental to the compression effect~\cite{lavoie13comp}, and because of this we will work in the separable limit here where the parameter $S\rightarrow\infty$.  In the first-order approximation, the bandwidth of the compressed pulse $\sigma_3^{(1)}$ is given by \begin{equation}\left(\sigma_3^{(1)}\right)^2=\frac{\sigma_1^2+\sigma_2^2}{1+16A^2\sigma_1^2\sigma_2^2}\stackrel{LCL}{\approx}\frac{1}{16A^2}\left(\frac{1}{\sigma_1^2}+\frac{1}{\sigma_2^2}\right).\end{equation}

For this discussion, we focus on the $q=1$ ($\sigma_1=\sigma_2=\sigma$) case, to gain insights into the regime experimentally explored in recent work by extrapolating to high-efficiency SFG.  In Fig.~\ref{bc}(a), we show numerical calculations of the upconversion probability as a function of the absolute coupling constant $\gamma$ for various chirps (and hence compression factors).  We see that the maximum possible efficiency is 88.7\%, regardless of the value of compression.  This represents a great potential improvement over lossy filtering techniques, but it is important to note that much higher coupling strengths require commensurate increases in escort power or material nonlinearity.

By Taylor-expanding the output temporal waveform $f_{3f}(t,t_h)$ about $\gamma$ and Fourier-transforming the result, we may express the spectrum as a summation. We numerically evaluate the effective width ${\sigma_3=\sqrt{\langle\omega_3^2\rangle-\langle\omega_3\rangle^2}}$ of the upconverted signal with no time delay ($\tau=0$) for various chirps $A$ and input bandwidths $\sigma_1$ and $\sigma_2$.  In Fig.~\ref{bc}(b), we show the ratio of this effective width to the first-order prediction as a function of the $q$ parameter and of the initial pulse length ratio ${q_0=(\sigma_2^2/\sigma_1^2)}$, representing the $q$ value before applying chirps.  In the low-$q$ limit, the high-order spectral width is identical to the expected value from a first-order calculation. In the case where the two temporal widths are exactly equal ($\sigma_1=\sigma_2$), the full-order spectrum after bandwidth compression is actually \emph{narrower} than first-order calculations predict.  This is because in this case the temporal waveform is flattened and thus has a larger full-width at half-maximum.  This effect is seen to persist near the diagonals of Fig.~\ref{bc}(b).  One can also see that the bandwidth is increased relative to the first-order calculation in the high-$q$ regime far from the diagonals; however, the relative increase in bandwidth is fairly small.

\begin{figure}[t!]
  \begin{center}
    \includegraphics[width=1\columnwidth]{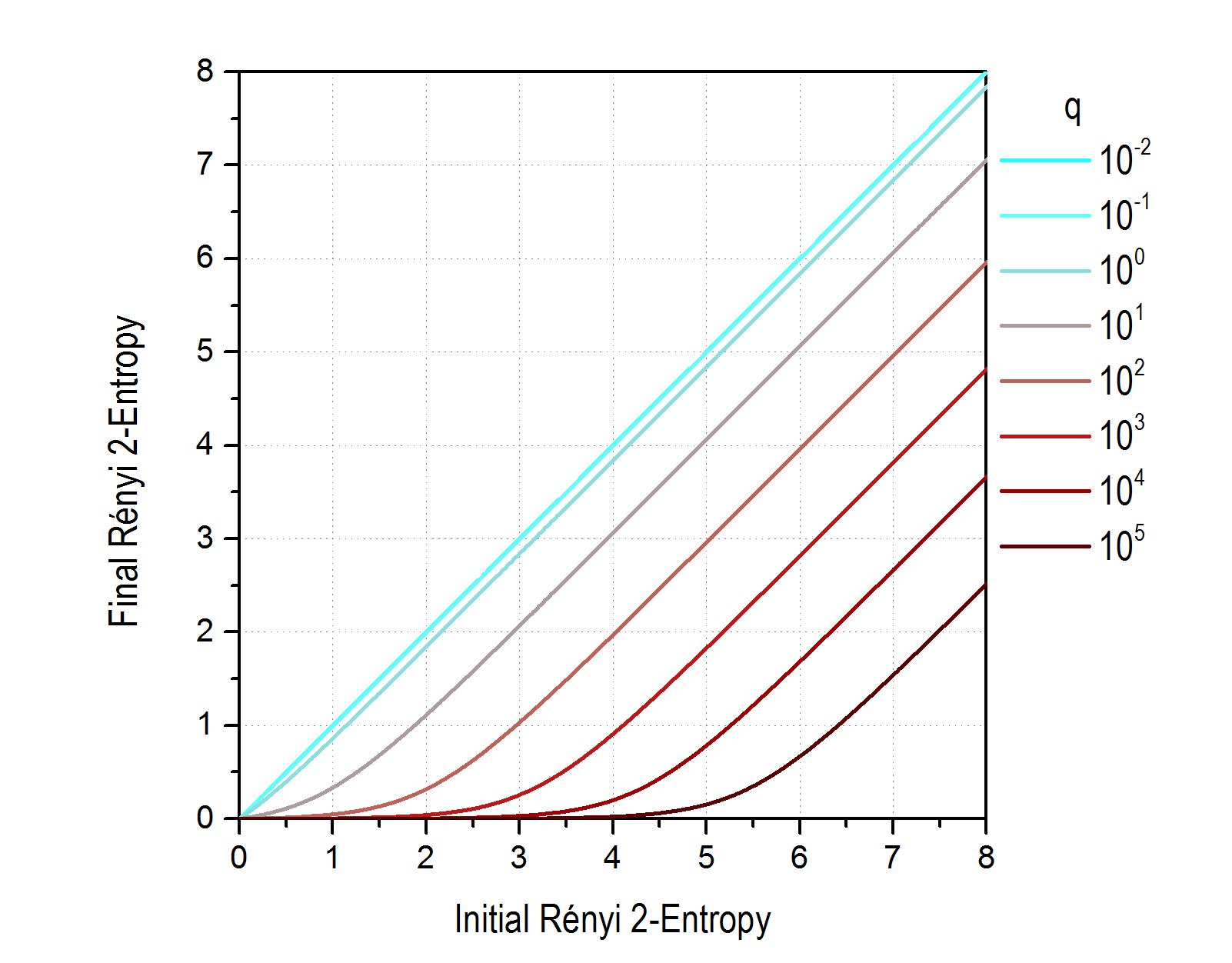}
  \end{center}
 \caption{\textbf{R\'{e}nyi 2-Entropy after upconversion at peak efficiency.} The R\'{e}nyi 2-entropy, a measure of entanglement, is shown for the upconverted subsystem as a function of the initial R\'{e}nyi 2-entropy for various values of $q$.  The bipartite energy-time entanglement is unaltered after post-selecting on successful upconversion if the pulse length ratio $q$ is small, i.e. in the time-lensing limit. However, if the input photon is of temporal length comparable to or longer than the strong escort pulse (high-$q$), the post-selection results in an effective loss of entanglement in addition to imperfect efficiency.}\label{Renyi}
\end{figure}

\section{Effectiveness with entanglement}

Entangled photons are critical to quantum technologies. Photon pairs with separable joint spectral amplitudes are desirable for numerous applications~\cite{bouwmeester1999observation,kim2002generation,branczyk2010optimized}, and applications such as multiplexing~\cite{jiang2006highly,christ2012exponentially,herbauts2013demonstration} require photon pairs with a multimode entangled structure. To retain these advantages, it is essential that the waveform manipulation process maintains entanglement.   In this section, we study the change in the degree of entanglement between the photons before and after upconversion.

We quantify the entanglement in the system through the R\'{e}nyi 2-entropy $\Upsilon(\rho)$~\cite{horodecki1996quantum}, defined for the density operator of a subsystem $\rho_{S}$ belonging to a bipartite pure state as \begin{equation}\Upsilon(\rho)=-\ln\Tr\rho_S^2.\end{equation} This measure quantifies the purity of the subsystem of a pure state; a high value indicates a high degree of entanglement and a correspondingly low purity for the individual subsystems. Given a pure bipartite state $\rho$ with a joint spectrum given by $F(\omega,\omega_h)$, the purity of the subsystem may be calculated based on either subsystem as \begin{equation}\Tr\rho_S^2=\int\dee\vec{\omega}\,F(\omega,\omega_h)F^*(\omega',\omega_h)F(\omega',\omega_h')F^*(\omega',\omega_h),\label{PurityJointSpec}\end{equation} where $\dee\vec{\omega}$ implies integration over all frequency variables. This expression may be adapted to a temporal representation through straightforward substitution.  For the joint spectrum of the input waveform given by Eq.~\eqref{inputspec}, the purity is found to be \begin{equation}\Tr\rho_{S,in}^2=\frac{S\sqrt{S^2+\sigma_1^2+\sigma_h^2}}{\sqrt{S^2+\sigma_1^2}\sqrt{S^2+\sigma_h^2}},\end{equation} which tends toward zero as $S$ approaches zero (maximal entanglement) and one as $S$ approaches infinity (separable).

Following the same method for the entanglement between the upconverted photon and the herald with $\tau=0$, we find \begin{widetext}\begin{equation}\Tr\rho_{S,up}^2=\frac{1}{\langle\hat{n}_3\rangle^2}\frac{1}{2\sqrt{2}}\sum_{m,n=1}^{\infty}\frac{(-1)^{m+n}S\sqrt{\frac{S^2+\sigma_1^2+\sigma_h^2}{2(1+mq)(1+nq)S^2(S^2+\sigma_1^2+\sigma_h^2)+(2+mq+nq)\sigma_1^2\sigma_h^2}}}{(2m)!(2n)!} p^{2(m+n)},\label{PurityUpconverted}\end{equation}\end{widetext} where $\langle\hat{n}_3\rangle$, $p$, and $q$ are as defined in Eqs.~(\ref{SumGen}-\ref{SumGenp}).  Examining this expression shows that the entanglement in the output is the same as that of the input in the limits of $S$ going to 0 or infinity (i.e., maximally entangled or fully separable).  In general, however, this is not the case.

Fig.~\ref{Renyi} compares the R\'{e}nyi entropy of the input photon pair to the R\'{e}nyi entropy between the upconverted photon and the herald with $p$ set to the optimal conversion efficiency, for a range of $q$ values, holding $\tau=0$.  It is seen that in the low-$q$ regime the R\'{e}nyi entropy, and hence entanglement, is effectively conserved through the SFG process, but not for high-$q$.  Note that our calculation applies for analysis of the upconverted pulse assuming Gaussian spectra and no time delay. For the calculations presented here, the entanglement of the final state is always lower than the initial.  However, for other spectral shapes, it could be that the upconverted subsystem actually has a higher degree of entanglement than the initial state if the upconversion process increases the number of significant Schmidt modes, akin to the Procrustean method of entanglement concentration~\cite{bennett1996concentrating}.

\section{Conclusion}

We have developed a theoretical treatment for high-efficiency sum-frequency generation between a single photon and a strong escort beam.  We have identified the parameter regime required for high-efficiency upconversion of the single photon for waveform manipulation.  This regime also coincides with the case where the photons remain Gaussian through upconversion and the first-order perturbative calculation provides an accurate description of the sum-frequency process.  We have found the conditions required to use SFG to implement a single-photon time lens and for bandwidth compression.  Finally, we have examined the effect of this process on two-photon energy-time entanglement, finding the regime where entanglement is conserved.  Nonlinear interactions mediated by shaped strong classical laser pulses provide a powerful platform for controlling the properties of quantum states of light.

\section*{Acknowledgements}

The authors thank A.\,M.~Bra\'{n}cyzk, A.\,A.~Burkov, C.\,M.~Herdman, J.~Lavoie, E.~Mart\'{i}n-Mart\'{i}nez, N.~Quesada, and  C.\,M.~Wilson for fruitful discussions.  We are grateful for financial support from the Natural Sciences and Engineering Research Council, the Canada Foundation for Innovation, Ontario Centres of Excellence, Industry Canada, the Canada Research Chairs Program, the Ontario Ministry of Training, Colleges, and Universities, and the Ontario Ministry of Research and Innovation.

\appendix

\section{Derivation of the sum-frequency generation unitary transformation}

We lay the groundwork for the various photon shaping techniques by following \cite{yang2008spontaneous} and defining our quantized electric field as propagating in the $\hat{z}$ direction, simplifying our operator representation to \begin{equation}\hat{E}(z,t)=\frac{i}{\sqrt{2\pi}}\int\dee k\sqrt{\frac{\hbar\omega_k}{2}}E_0\hat{a}_{\omega_k}e^{ikz-i\omega_kt}+h.c.\end{equation} We approximate the transverse profile of the electric field to be uniform within an area $\mathcal{A}$ (effectively the active area in the nonlinear medium) and zero elsewhere. We determine the constant $E_0$ by demanding that the energy in a single mode $k$ is $\hbar\omega_k$ with the approximations that the permittivity is $\epsilon\approx\epsilon_0n(\omega)$ and that the group and phase velocities are equal ($\frac{\dee n}{\dee\omega}\ll\frac{n}{\omega}$), finding that \begin{equation}\iint_\mathcal{A}\dee x\dee y\,\frac{\epsilon_0n^2(\omega_k)E_0^2}{2}\hbar\omega_k=\frac{\epsilon_0n^2(\omega_k)E_0^2}{2}\mathcal{A}\hbar\omega_k=\hbar\omega_k.\end{equation} We re-express the integral in terms of frequency rather than wavenumber, as in Section V of \cite{yang2008spontaneous} with the additional approximation $\frac{\dee n}{\dee\omega}\ll\frac{n}{\omega}$, as \begin{equation}\int\dee k\,\hat{a}_{\omega_k}=\int\dee\omega\,\sqrt{\frac{\dee k}{\dee\omega}}\hat{a}_\omega\approx\int\dee\omega\,\sqrt{\frac{n(\omega)}{c}}\hat{a}_\omega.\end{equation} We split our field operator into its positive- and negative-frequency components, ${\hat{E}(z,t)=\hat{E}^{(+)}(z,t)+\hat{E}^{(-)}(z,t)}$, where \begin{align}E^{(+)}(z,t)&=i\int\dee\omega\,\sqrt{\frac{\hbar k(\omega)}{4\pi\epsilon_0\mathcal{A}}}\hat{a}_\omega e^{-i\omega t+ikz}\\ E^{(-)}(z,t)&=-i\int\dee\omega\,\sqrt{\frac{\hbar k(\omega)}{4\pi\epsilon_0\mathcal{A}}}\hat{a}^\dag_\omega e^{i\omega t-ikz},\end{align} where $k(\omega)=n(\omega)\omega/c$.

We assume that the three fields involved in the upconversion process (input signal, strong escort, and generated signal) are completely non-degenerate and thus may be viewed as occupying three independent modes, numbered 1-3 and with annihilation operators defined as $\hat{a}$, $\hat{g}$, and $\hat{c}$ respectively. We then define the interaction Hamiltonian $\hat{H}_I(t)$  for the upconversion process as~\cite{kumar1990quantum} \begin{equation}\hat{H}_I(t)=-\frac{\epsilon_0}{3}\chi^{(2)}\int_V\dee \vec{r}\left(E_1^{(-)}E_2^{(-)}E_3^{(+)}+\mathrm{h.c.}\right)\end{equation} After integrating over the transverse profile, we rewrite the Hamiltonian as \begin{widetext}\begin{equation}\hat{H}_I(t)=-\frac{\epsilon_0}{3}\chi^{(2)}\left(\frac{\hbar}{4\pi\epsilon_0\mathcal{A}}\right)^{\frac{3}{2}}\mathcal{A}\iiiint\dee\omega_1\dee\omega_2\dee\omega_3\dee z\sqrt{k_1k_2k_3}\left[i\hat{a}_{\omega_1}\hat{g}_{\omega_2}\hat{c}_{\omega_3}^\dag e^{-i(\omega_1+\omega_2-\omega_3)t}e^{i(k_1+k_2-k_3)z}+\mathrm{h.c.}\right],\label{HamiltonianAppA}\end{equation} where $k_i$ is a function of $\omega_i$.

The initial state evolves according the unitary transformation $\hat{U}\equiv\TO\exp[-\frac{i}{\hbar}\int_{t_0}^{t_f}\dee t\hat{H}_I(t)]$, where $\TO$ is the time ordering operator~\cite{branczyk2010non,branczyk2011time,christ13theory,quesada2014effects,quesada2014time}.  In general, time ordering necessitates corrections to the Taylor expansion of this transformation. However, in Section II of the main text, we showed that these corrections vanish in the limit of perfect phasematching; we neglect these corrections for the remainder of the discussion. We additionally assume that the interaction Hamiltonian is zero at before $t_0$ and after $t_f$, allowing us to extend the limits of integration in the unitary transformation to infinity and obtain the delta function $2\pi\delta(\omega_3-\omega_1-\omega_2)$.  While we aim to study spectral waveforms that are not strictly monochromatic, we make the assumption that our waveforms are relatively narrowband, such that $\sqrt{k_1k_2k_3}$ is approximately the constant $\sqrt{k_{01}k_{02}k_{03}}$ (where $k_{0i}=k(\omega_{0i})\,$) and may be taken out of the integral.  We then take the $z$ integral over the crystal length $L$ to find \begin{align}\int_{-L/2}^{L/2}\dee z\,e^{i(k_1+k_2-k_3)z}&=L\,\textrm{sinc}\left[\frac{1}{2}L(k_1+k_2-k_3)\right]\nonumber\\&=L\,\Phi(\omega_1,\omega_2,\omega_3),\end{align} where $\Phi(\omega_1,\omega_2,\omega_3)$ is the phasematching function.

We can rewrite the evolution of the system as defined by the simplified unitary transformation \begin{equation}\hat{U}_{wm}=\exp\left\{i\left(\frac{\hbar k_{01}k_{02}k_{03}}{64\pi^3\epsilon_0\mathcal{A}}\right)^{\frac{1}{2}}\frac{2\pi\chi^{(2)}L}{3}\iiint\dee\omega_1\dee\omega_2\dee\omega_3\, \left[i\hat{a}_{\omega_1}\hat{g}_{\omega_2}\hat{c}_{\omega_3}^\dag\Phi(\omega_1,\omega_2,\omega_3)+h.c. \right]\delta(\omega_1+\omega_2-\omega_3)\right\}.\label{unitary}\end{equation} Since the second mode is a strong coherent state, we may make the approximation such that the operators may be treated as constant (the non-depleted pump approximation), where we define $\hat{g}_{\omega_2}\approx \sqrt{N_e} G(\omega_2)$ where $N_e$ represents the number of photons in the escort pulse and $G(\omega_2)$ is its complex normalized spectrum. We also absorb the factor of $i$ into $G(\omega_2)$. We re-express the transformation of Eq.~\eqref{unitary} as \begin{equation}\hat{U}_{SFG}=\exp\left\{i\left(\frac{\hbar k_{01}k_{02}k_{03}}{16\pi\epsilon_0\mathcal{A}}\right)^{\frac{1}{2}}\frac{\chi^{(2)}L\sqrt{N_e}}{3} \iint\dee\omega_1\dee\omega_3\,\left[\hat{a}_{\omega_1}\hat{c}_{\omega_3}^\dag G(\omega_3-\omega_1)\Phi(\omega_1,\omega_3-\omega_1,\omega_3)+h.c.\right]\right\}.\label{unitary2}\end{equation}\end{widetext} By collecting all constant terms, we define the \emph{absolute coupling constant}  \begin{equation}\gamma=\left(\frac{\hbar k_{01}k_{02}k_{03}}{16\pi\epsilon_0\mathcal{A}}\right)^{\frac{1}{2}}\frac{\chi^{(2)}L\sqrt{N_e}}{3},\end{equation} which has dimensions of square-root-time, such that ${\int\dee\omega\,\gamma^2\,|G(\omega)|^2}$ is dimensionless. By taking the Taylor expansion about $\gamma=0$, we find the unitary transformation defined in Eq.~\eqref{unitary4} of the main text.

\section{Time-domain expression for sum-frequency generation}

With an input state and evolution unitary transformation written as in Eq.~\eqref{stateinit} and Eq.~\eqref{unitary4} of the main text respectively, the final state is given as an infinite sum of terms in the frequency domain. In this section, we show the transformations from the frequency to the time domain necessary to obtain a general solution, starting from Eq.~\eqref{unitary4} and ending with Eqs.~(\ref{f1f}-\ref{f3f}) of the main text.  To do so, we use the Fourier transform to switch between time and angular frequency representations, in a form designed to maintain the normalization of the integrated photon number, \begin{equation}f(t)=\mathscr{F}^{-1}\left[F(\omega)\right]=\frac{1}{\sqrt{2\pi}}\int_{-\infty}^\infty\dee\omega F(\omega)e^{i\omega t},\label{fourier}\end{equation} and define the escort pulse temporal representation ${g(t)=\mathscr{F}^{-1}\left[G(\omega_2)\right]}$ and the input joint temporal waveform ${f_i(t,t_h)=\mathscr{F}^{-1}\left[F_i(\omega_1,\omega_h)\right]}$ via a straightforward two-dimensional generalization of Eq.~\eqref{fourier}.

As the photon number must be conserved, the action of SFG will be to couple modes 1 and 3 (see Fig.~\ref{concept}(a) of the main text) as \begin{widetext}\begin{align}\ket{\psi_f(t)}=\frac{1}{2\pi}\sum_{k=0}^\infty\int\dee\omega_h\left[\int\dee\omega_1F^{(2k)}(\omega_1,\omega_h)\ket{\omega_1}_1\ket{0}_3\ket{\omega_h}_h +\int\dee\omega_3F^{(2k+1)}(\omega_3,\omega_h)\ket{0}_1\ket{\omega_3}_3\ket{\omega_h}_h\right],\label{psiF}\end{align} with odd-order terms representing sum-frequency conversion from mode 1 to mode 3 and even-order terms representing difference-frequency conversion from mode 3 back to mode 1.  Noting that ${F^{(0)}(\omega_1,\omega_h)=F_i(\omega_1,\omega_h)}$, the first-order term may be directly calculated as \begin{equation}F^{(1)}(\omega_3',\omega_h)=i\gamma\int\dee\omega_1F^{(0)}(\omega_1,\omega_h)G(\omega_3'-\omega_1),\end{equation} and the convolution theorem may be used to express this term in the time domain as simply ${f^{(1)}(t,t_h)=i\sqrt{2\pi}\gamma f^{(0)}(t,t_h)g(t)}$.

Higher-order contributions may be found recursively through repeated convolution-like integrals over the signal frequency with the escort beam, as \begin{align}F^{(k+2)}_{even}(\omega_1'',\omega_h)&=\frac{-\gamma^2}{(k+2)(k+1)}\iint^{\infty}_{-\infty}\dee\omega_1\dee\omega_3'F^{(k)}(\omega_1,\omega_h) G(\omega_3'-\omega_1) G^*(\omega_3'-\omega_1'')\nonumber\\ F^{(k+2)}_{odd}(\omega_3'',\omega_h)&=\frac{-\gamma^2}{(k+2)(k+1)}\iint^{\infty}_{-\infty}\dee\omega_1'\dee\omega_3F^{(k)}(\omega_3,\omega_h) G^*(\omega_3-\omega_1') G(\omega_3''-\omega_1').\label{freqrecur}\end{align} These integrals are highly similar to convolution, but differ in a subtle fashion, as the term $G^*$ appears as the function $G^*(\omega_3-\omega_1)$ rather than $G^*(\omega_1-\omega_3)$.  The double-conversion in question for the upconverted mode is of the form \begin{equation}F^{(k+2)}(\omega'')=\iint_{-\infty}^{\infty}\dee\omega\dee\omega'F^{(k)}(\omega) G^*(\omega-\omega') G(\omega''-\omega'),\end{equation} where we have neglected the herald frequency for clarity. Were this describing the third-order upconversion, for example, $F^{(1)}(\omega)$ would represent the first-order upconversion, $\omega$ the frequency of the first-order blue photon created, and $\omega'$ the frequency of the second-order red photon created. Letting $\mathscr{F}^{-1}\left[ G(\omega)\right]= g(t)$, $\mathscr{F}^{-1}\left[F^{(k)}(\omega)\right]=f^{(k)}(t)$ and assuming that all integrals extend to infinity implicitly, we find that \begin{align}F^{(k+2)}(\omega'')&=\iint\dee\omega\dee\omega' F^{(k)}(\omega)\left[\frac{1}{\sqrt{2\pi}}\int\dee t g(t)e^{-i(\omega-\omega')t}\right]^* G(\omega''-\omega')\nonumber\\ &=\iint\dee\omega\dee\omega' F^{(k)}(\omega)\left[\frac{1}{\sqrt{2\pi}}\int\dee t g^*(t)e^{i(\omega-\omega')t}\right] G(\omega''-\omega')\nonumber\\ &=\frac{1}{\sqrt{2\pi}}\int\dee t \left[\int\dee\omega F^{(k)}(\omega)e^{i\omega t}\right] g^*(t) \left[\int\dee\omega' G(\omega''-\omega')e^{-i\omega't}\right]\nonumber\\ &=\sqrt{2\pi}\int\dee t \left[\frac{1}{\sqrt{2\pi}}\int\dee\omega F^{(k)}(\omega)e^{i\omega t}\right] g^*(t) \left[\frac{1}{\sqrt{2\pi}}\int\dee\omega' G(\omega''-\omega')e^{i(\omega''-\omega')t}\right]e^{-i\omega''t}\nonumber\\ &=\sqrt{2\pi}\int\dee t f^{(k)}(t) g^*(t) g(t)e^{-i\omega''t}\nonumber\\&=2\pi\mathscr{F}^{-1}\left[f^{(k)}(t) g^*(t) g(t)\right].\end{align}

By also transforming the herald to the time domain, we may re-express the recursion relation in the time domain as \begin{equation}f^{(k+2)}(t,t_h)=\frac{-2\pi\gamma^2}{(k+1)(k+2)}f^{(k)}(t,t_h)\left| g(t)\right|^2.\label{RecurTime}\end{equation} Using the previously defined $f^{(0)}(t,t_h)$ and $f^{(1)}(t,t_h)$ as base cases, the final temporal waveforms are found to be \begin{align}f_{1f}(t,t_h)&=\sum_{\mathrm{even}\,k}^{\infty}\frac{(i\sqrt{2\pi}\gamma)^k}{k!}f_i(t,t_h)|g(t)|^k=f_i(t,t_h)\cos\left[\sqrt{2\pi}\gamma|g(t)|\right], \\f_{3f}(t,t_h)&=\sum_{\mathrm{odd}\,k}^{\infty}\frac{(i\sqrt{2\pi}\gamma)^k}{k!}f_i(t,t_h)\frac{g(t)}{|g(t)|}|g(t)|^k=f_i(t,t_h)\frac{g(t)}{|g(t)|}\sin\left[\sqrt{2\pi}\gamma|g(t)|\right].\end{align}

For the specific state and escort profile defined in Eqs.~\eqref{inputspec} and \eqref{escortspec} respectively, the final temporal waveforms are found from Eqs.~\eqref{f1f} and \eqref{f3f} with $\sigma_{in}^2=S^2+\sigma_1^2+\sigma_h^2$ and $\zeta_j=1-4iA_j\sigma_j^2$ as \begin{align} f_{1f}(t,t_h)=&\left(\frac{2S\sigma_1\sigma_h\sigma_{in}}{\pi\left[\zeta_1(S^2+\sigma_h^2)+\sigma_1^2\right]}\right)^\frac{1}{2} e^{-\frac{\sigma_{in}^2\left[\sigma_1^2\sigma_h^2(t-t_h)^2+S^2(\sigma_1^2t^2+\sigma_h^2t_h^2)\right]-16A_1^2\sigma_1^4\sigma_h^2S^2(S^2+\sigma_h^2)t_h^2}{\sigma_{in}^4+16A_1^2\sigma_1^4(S^2+\sigma_h^2)^2}}\nonumber\\&\times e^{-4i\left[\frac{A_1\sigma_1^4(S^2t+\sigma_h(t-t_h))^2}{\sigma_{in}^4+16A_1^2\sigma_1^4(S^2+\sigma_h^2)^2}\right]}e^{i\omega_{01}t}e^{i\omega_{0h}t_h}\cos\left[\left(\frac{8\pi\sigma_2^2}{|\zeta_2|^2}\right)^\frac{1}{4}{\gamma}e^{-\frac{\sigma_2^2(t+\tau)^2}{|\zeta_2|^2}}\right] \\f_{3f}(t,t_h)=&i\left(\frac{2S\sigma_1\sigma_h\sigma_{in}}{\pi\left[\zeta_1(S^2+\sigma_h^2)+\sigma_1^2\right]}\frac{\zeta_2^*}{|\zeta_2|}\right)^\frac{1}{2} e^{-\frac{\sigma_{in}^2\left[\sigma_1^2\sigma_h^2(t-t_h)^2+S^2(\sigma_1^2t^2+\sigma_h^2t_h^2)\right]-16A_1^2\sigma_1^4\sigma_h^2S^2(S^2+\sigma_h^2)t_h^2}{\sigma_{in}^4+16A_1^2\sigma_1^4(S^2+\sigma_h^2)^2}}\nonumber\\&\times e^{-4i\left[\frac{A_1\sigma_1^4(S^2t+\sigma_h(t-t_h))^2}{\sigma_{in}^4+16A_1^2\sigma_1^4(S^2+\sigma_h^2)^2}+\frac{A_2\sigma_2^4(t+\tau)^2}{|\zeta_2|^2}\right]}e^{i(\omega_{01}+\omega_{02})t}e^{i\omega_{0h}t_h}\sin\left[\left(\frac{8\pi\sigma_2^2}{|\zeta_2|^2}\right)^\frac{1}{4}{\gamma}e^{-\frac{\sigma_2^2(t+\tau)^2}{|\zeta_2|^2}}\right].\end{align}\end{widetext}

%


\end{document}